\documentclass[%
aip,apl,twocolumn,reprint,amsmath,amssymb,showpacs,superscriptaddress,citeautoscript]{revtex4-2}

\usepackage{epsfig}
\usepackage{amsmath}
\usepackage{graphics}
\usepackage{bm}
\usepackage{makecell}
\usepackage{float}
\usepackage{multirow}
\usepackage{dcolumn}
\usepackage{hyperref}
\usepackage{physics}
\usepackage{xr}
\makeatletter
\newcommand*{\addFileDependency}[1]{
  \typeout{(#1)}
  \@addtofilelist{#1}
  \IfFileExists{#1}{}{\typeout{No file #1.}}
}
\makeatother

\newcommand*{\myexternaldocument}[1]{%
    \externaldocument{#1}%
    \addFileDependency{#1.tex}%
    \addFileDependency{#1.aux}%
}
\myexternaldocument{Supplement}

\hypersetup{
	colorlinks=true,
	linkcolor=blue,
	filecolor=gray,
	urlcolor=blue,
	citecolor=blue,
}
\begin{document}
\preprint{APS/123-QED}

\title{Carbon in GaN as a nonradiative recombination center}

\author{Fangzhou Zhao}
\affiliation{Materials Department, University of California, Santa Barbara, California 93106-5050, USA 
}
\affiliation{Theory Department, Max Planck Institute for Structure and Dynamics of Matter and Center for Free-Electron Laser Science, 22761 Hamburg, Germany}
\date{\today}
\author{Hongyi Guan}
\affiliation{Materials Department, University of California, Santa Barbara, California 93106-5050, USA 
}
\author{Mark E. Turiansky}
\altaffiliation[Present Address: ]{US Naval Research Laboratory, 4555 Overlook Avenue SW, Washington, DC 20375, USA}
\affiliation{Materials Department, University of California, Santa Barbara, California 93106-5050, USA 
}
\author{Chris G. Van de Walle}
\affiliation{Materials Department, University of California, Santa Barbara, California 93106-5050, USA 
}

\begin{abstract}
Trap-assisted nonradiative recombination has been shown to limit the efficiency of optoelectronic devices. 
While substitutional carbon ($\mathrm{C_N}$) has been suggested to be a nonradiative recombination center in GaN devices, a complete recombination cycle including the two charge-state transition levels has not been previously described. 
In this work, we investigate the trap-assisted recombination process due to $\mathrm{C_N}$ in GaN, including multiphonon emission (MPE), radiative recombination, trap-assisted Auger-Meitner (TAAM) recombination, as well as thermal emission of holes. 
Our study shows the key role of TAAM processes at the high carrier densities relevant for devices.
We also reveal the carrier-density regimes where thermal emission and radiative recombination are expected to play an observable role. 
Our results highlight that carbon concentrations exceeding $\sim$10$^{17}$~cm$^{-3}$ can have a noticeable impact on device efficiency, not just in GaN active layers but also in InGaN and AlGaN.
Our comprehensive formalism not only offers detailed results for carbon but provides a general framework for assessing the multiple processes that participate in trap-assisted recombination in semiconductors. 
\end{abstract}

\maketitle

Gallium nitride (GaN)-based light-emitting diodes (LEDs) have revolutionized illumination technology, enabling highly energy-efficient solid-state lighting~\cite{10.1063/1.111832,Feezell_2018}, lasers~\cite{HARDY2011408},
and extending into the ultraviolet (UV) spectrum~\cite{Kneissl2019}.
Internal quantum efficiencies are still limited by point defects, however~\cite{pssa.202100727}, and the microscopic origin of these detrimental defects is still under debate.
Since efficiency losses are universally observed, native point defects (such as cation or nitrogen vacancies) are commonly blamed.  Here we examine another candidate, which is present in all nitride materials grown by metal-organic chemical vapor deposition (MOCVD): unintentional carbon impurities.
The most common form in which carbon is incorporated in nitride active layers (which are typically uninentionally $n$-type) is as a  substitutional acceptor on the nitrogen site, C$_{\rm N}$, with a defect level approximately 1~eV above the valence-band maximum (VBM).  Because this level is far from midgap, nonradiative recombination due to C$_{\rm N}$ was not seriously considered in the past; this assessment was based on a traditional view of trap-assisted nonradiative recombination, going back to the seminal work of Shockley and Read~\cite{SR} and Hall~\cite{Hall}.  However, recent experimental work indicates that carbon does impact nonradiative recombination~\cite{kojima2019roles,kojima_2024}, highlighting the need for understanding the underlying mechanisms.

\begin{figure}[!b]
	\centering
	\includegraphics[width=\linewidth]{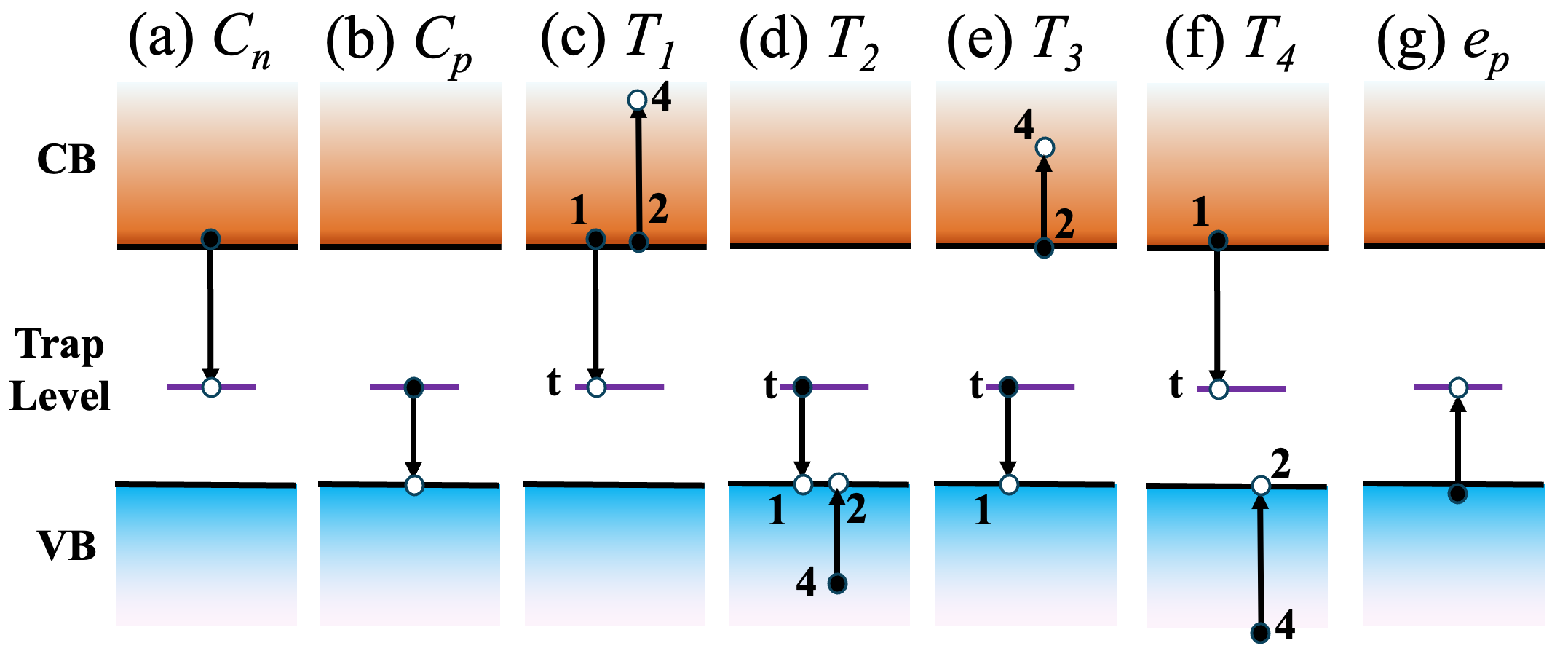}
	\caption{Schematic diagram of trap-assisted recombination processes considered in this work. (a),(b): electron and hole capture, either radiatively or nonradiatively via MPE; (c)-(f): TAAM processes labeled by $T_1$ through $T_4$; (g): hole  emission process. Solid and hollow circles denote electrons and holes, respectively.}
	\label{SRH_TAAM_TE}
\end{figure}

Shockley-Read-Hall (SRH) recombination via multiphonon emission (MPE) \cite{Abakumov1991} is conventionally used to describe the loss mechanism at deep centers \cite{PhysRevB.90.075202}. 
In an MPE process [Figs.~\ref{SRH_TAAM_TE}(a),(b)], electrons and holes are captured at an in-gap trap level and the recombination energy is transferred to lattice vibrations. 
Various defects have been proposed to explain efficiency loss in InGaN LEDs based on recombination by MPE~\cite{Dreyer_2016,Shen_2017}; 
however, the MPE capture rate decreases exponentially as the energy difference between the trap level and the band edge increases~\cite{PhysRevB.15.989,PhysRevB.90.075202,DiBartolo1991}.
In the case of carbon, the position of the defect level allows for fast hole capture, but electron capture would be very slow in most visible and UV devices.
Since the rate of the overall SRH recombination cycle is limited by the slower of the electron or hole capture process~\cite{Dreyer_2016}, an assessment based solely on MPE would conclude that carbon is ineffective as a nonradiative center.
In fact, this conclusion would apply to {\it any} defect in a semiconductor with bandgap larger than about 2.6~eV, since the trap level would always be too distant from either the VBM or the conduction-band minimum (CBM)~\cite{PhysRevB.93.201304,PhysRevB.15.989,Abakumov1991,Shen_2017}. 

To explain the observed recombination in wider-gap devices, additional processes therefore need to be invoked.
Radiative recombination can occur [and is included in Figs.~\ref{SRH_TAAM_TE}(a),(b)], but as we will see it is generally too slow to impact device efficiency.
Excited states of the defect could play a role~\cite{PhysRevB.93.201304}, but C$_{\rm N}$ does not possess such excited states.
The involvement of another recombination process has recently been proposed: trap-assisted Auger-Meitner (TAAM) recombination \cite{pssa.202100727,10.1063/1.5096773,PhysRevLett.131.056402}. 
As illustrated in Figs.~\ref{SRH_TAAM_TE}(c)-(f), TAAM involves two free carriers, one which is captured at a trap state, and a second one that absorbs the recombination energy by being excited to a higher-energy state~\cite{Abakumov1991,PhysRevLett.131.056402}. 
In recent work we demonstrated, based on first-principles calculations for the benchmark case of a calcium impurity, that TAAM recombination dominates over MPE in InGaN with bandgaps above 2.5 eV \cite{PhysRevLett.131.056402}. 
We also showed that, unlike MPE, the TAAM rate does not  strongly depend on the bandgap. 

Here we present a comprehensive first-principles study of trap-assisted recombination related to $\mathrm{C_N}$ in wurtzite GaN.
Thermodynamic transition levels for $\mathrm{C_N}$ have previously been reported~\cite{10.1063/1.3492841,PhysRevB.89.035204,10.1063/1.5047808,PhysRevB.90.235203}.
We calculate capture coefficients based on radiative, MPE and TAAM processes for electron and hole capture at the (+/0) and (0/$-$) levels. Since the (+/0) levels occurs quite close to the valence band, thermal emission from this level~\cite{PhysRevB.90.235203} must be considered, an issue that has not been addressed in previous first-principles studies of SRH recombination.  
Taking hole emission into account in the overall recombination cycle allows us to identify conditions where emission processes are important.

C$_{\rm N}$ impurities are already known to be a source of yellow luminescence~\cite{10.1063/1.3492841,https://doi.org/10.1002/pssb.202200488}, but we find that trap-assisted radiative recombination does not noticeably impact device efficiency.  
In contrast, at carrier densities above 5$\times$$10^{16}$~$\mathrm{cm^{-3}}$, TAAM recombination 
dominates electron capture 
and can lead to efficiency loss in GaN-based LEDs.
Since the TAAM process is relatively insensitive to transition energy, this conclusion also applies to InGaN- and AlGaN-based devices. 
We find that thermal emission from the (+/0) level suppresses the total recombination rate, but only when the injected carrier density is smaller than $\sim$7$\times$10$^{16}$~cm$^{-3}$.


The processes relevant for trap-assisted recombination, for the example of a single trap level in the bandgap, are summarized in Fig.~\ref{SRH_TAAM_TE}:
radiative or MPE-assisted nonradiative capture [(a)-(b)], TAAM recombination [(c)-(f)], and thermal emission (g). 
If the trap state is the $q$/($q-1$) level,  
the initial state for electron capture (a) is the $q$ charge state with density $N^q$, and the initial state for hole capture (b) the $q-1$ charge state with density $N^{q-1}$.  
TAAM processes include capture of a free electron, transferring energy to another free electron [process $T_1$ (c)] or hole [$T_4$ (f)]; or the capture of a free hole, accompanied by excitation of another free electron [$T_3$ (d)] or hole [$T_2$ (c)]  \cite{Abakumov1991}. TAAM recombination rates are second order in the carrier densities as two free carriers are involved.
MPE capture coefficients $C_n$ and $C_p$ have units of $\rm cm^3 s^{-1}$, while TAAM coefficients have units of $\rm cm^6 s^{-1}$.

\begin{figure}[!t]
	\centering
	\includegraphics[width=\linewidth]{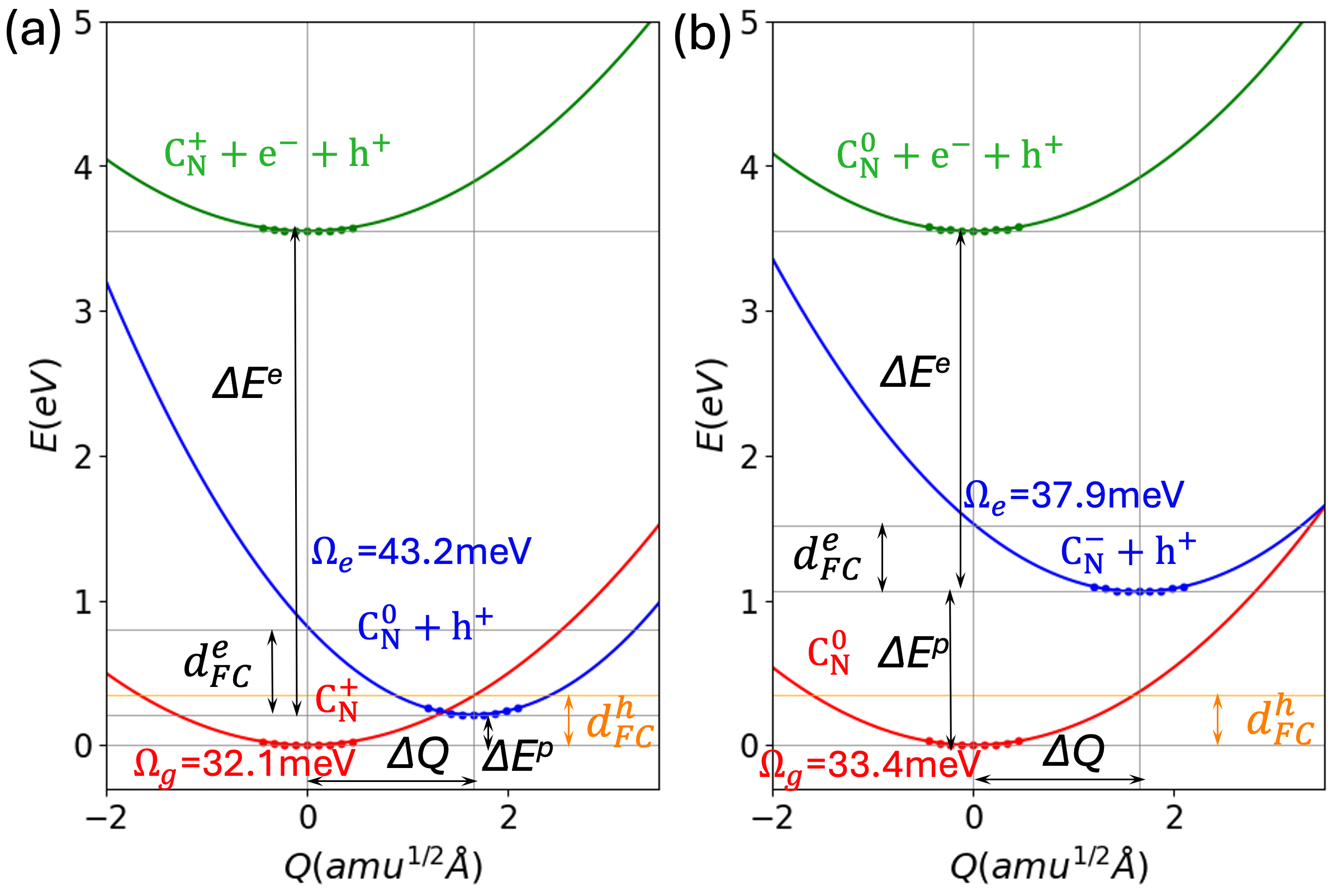}
	\caption{Configuration coordinate diagrams for $\mathrm{C_N}$ in GaN for carrier capture at the (a) (+/0) and (b) (0/$-$) level.  Symbols denote calculated values; solid lines are parabolic fits. 
}
	\label{CCD}
\end{figure}

MPE can be described in a configuration coordinate diagram (CCD) (see Fig.~\ref{CCD}).
Capture coefficients are evaluated using the formalism developed in Ref.~\onlinecite{PhysRevB.90.075202}, as implemented in the \texttt{Nonrad} code \cite{TURIANSKY2021108056}.
Trap-assisted radiative recombination coefficients are calculated according to Ref.~\onlinecite{dreyer2020radiative}.
For TAAM 
we use the formalism developed in Ref.~\onlinecite{PhysRevLett.131.056402}. 
Expressions for the four TAAM coefficients 
are included in Sec. S1
of the Supplementary Material (SM).
Since C$_{\rm N}$ has two levels in the gap, 
we specify the level that the coefficients refer to using a superscript $q$ on the $C$ and $T$ coefficients, 
where $q = +$, $0$, or $-$ indicates the initial charge state.
$C_{n,{\rm MPE}}^{q}$ and $C_{p,{\rm MPE}}^{q}$ are thus the MPE electron and hole capture coefficients at a trap with charge state $q$ in the initial state, and $C_{n,{\rm rad}}^{q}$ and $C_{p,{\rm rad}}^{q}$ are the corresponding radiative electron and hole capture coefficients. 
Similarly, we denote $T_1^q, T_2^{q-1}, T_3^{q-1}, T_4^q$ as the TAAM coefficient at level $q/(q-1)$.

Describing carrier capture at charged traps requires introducing a Sommerfeld parameter, $s(T)$, 
which describes the modification of the free-carrier wavefunction near a point charge
\cite{TURIANSKY2021108056,https://doi.org/10.1002/pssb.2220780222, dreyer2020radiative}.
This factor
enhances (suppresses) the calculated rate for an attractive (repulsive) center
and is evaluated using formulas in Ref.~\onlinecite{Turiansky_2024}.
We use a subscript $h$ or $e$ to indicate hole or electron capture, and a superscript $a$ or $r$ to indicate an attractive or repulsive interaction.
For the MPE and radiative capture coefficients, $C_n^+$ has a factor $s_e^a$ and $C_p^{-}$ has a factor $s_h^a$. 
For TAAM, we have factors $(s_e^a)^2$ for $T_1^+$, $(s_h^a)^2$ for $T_2^-$, $s_h^a s_e^r$ for $T_3^-$, and $s_e^a s_h^r$ for $T_4^+$. 
\footnote{We are assuming here that the carriers interact independently with the charged center, leading to a product of two Sommerfeld parameters. Since carriers can screen the center, some modification might be needed.  E.g., in the case of the $T_1^+$ process, the actual factor would be between $(s_{e}^a)$ (full screening) and $(s_{e}^a)^2$ (no screening). }

Semiconductor devices operate at finite temperature. 
The temperature dependence of MPE rates is significant and fully accounted for in our formalism~\cite{PhysRevB.90.075202}; for TAAM~\cite{PhysRevLett.131.056402} and radiative~\cite{dreyer2020radiative} recombination, we found that the temperature dependence is weak.
In addition, carriers in trap states can acquire sufficient thermal energy to be emitted back into the conduction or valence band, 
a process expected to be particularly relevant for trap levels that are close to a band edge. 
For C$_{\rm N}$ in GaN, only the (+/0) charge-state transition level is close enough to the band edge to lead to hole emission near room temperature. 
Using the principle of detailed balance~\cite{SR, 10.1063/1.5047808}, 
the hole thermal emission coefficient $e_p$ (units s$^{-1}$) can be calculated as~\cite{10.1063/1.1663719,10.1063/1.5047808}:
\begin{equation}
    e_p = C_pN_v\exp\qty(-\Delta E/kT),  \label{e_p} 
\end{equation}
where $\Delta E$ is the energy difference between the trap state and the band edge and 
$N_v$ is the effective density of states for the valence band (in which we use the density-of-states hole mass $m_h$=1.25~$m_0$ \cite{hole/mass}). 


To determine the overall trap-assisted recombination rate, we need to take into account the two thermodynamic transition levels of $\mathrm{C_N}$: (+/0) and (0/$-$) \cite{PhysRevB.89.035204,PhysRevB.90.235203,10.1063/5.0041608}. 
The electron (hole) capture rate for the combined MPE and radiative processes at charge-state transition level $q/(q-1)$ is $R_n^q = N^{q} n C_n^q $ ($R_p^{q-1} = N^{q-1} p C_p ^{q-1} $),  where 
we have defined $C^q_{p/n} = C^q_{p/n,{\rm MPE}} + C^q_{p/n,{\rm rad}}$ (Sec. S2
of the SM), 
and $n$ and $p$ are the electron  and hole carrier densities.
Similarly, the rates for TAAM recombination processes are
$R_{n,1}^q = N^{q} T_1^q n^2$, $R_{p,2}^{q-1} =N^{q-1} T_2^{q-1} p^2$, $R_{p,3}^{q-1} =N^{q-1} T_3^{q-1} np$ and $R_{n,4}^q =N^{q} T_4^q np$. 
The thermal emission rate from the $(+/0)$ level is $e_p N^+$.
For completeness, we also include electron thermal emission from the $(0/-)$ level, with a coefficient $e_n$ and a rate $e_n N^-$.


All quantities are calculated from first principles using density function theory with the Heyd-Scuseria-Ernzerhof (HSE) hybrid functional~\cite{heyd2003hybrid, heyd2006hybrid} implemented in the \texttt{QUANTUM ESPRESSO} package~\cite{Giannozzi_2009}.
Details of the computations are included in Sec. S3 
of the SM. 
The $\mathrm{C_N}$ impurity is calculated according to the methodology of Ref.~\onlinecite{RevModPhys.86.253},
using 96-atom supercells (with a 2$\times$2$\times$2 Monkhorst-Pack $\vb{k}$-point mesh \cite{PhysRevB.13.5188}) and 360-atom supercells ($\Gamma$ point only).  
Spin polarization is included and finite-size corrections for charged systems are applied \cite{freysoldt2009fully, freysoldt2011electrostatic}.
MPE coefficients are calculated using the \texttt{Nonrad} code \cite{TURIANSKY2021108056} and the 96-atom supercell. 
TAAM coefficients are calculated according to Ref.~\onlinecite{PhysRevLett.131.056402} and found to be converged for the 360-atom supercell. 
TAAM calculations take phonon assistance into account [see Eqs.~(S1)-(S4) in the SM].
The calculated Huang-Rhys (HR) factors for both electron and hole capture process at both trap levels are large, allowing us to include broadening by using Gaussian functions with a width of 0.22~eV.
 

Our calculated formation energies for $\rm C_N$ under Ga-rich conditions in the +, 0, and $-$ charge states are shown in Sec. S3
of the SM. 
Two thermodynamic transition levels occur within the gap: (+/0) at 0.21~eV above the VBM and (0/$-$) at 1.06~eV above the VBM, in agreement with previous computational 
\cite{10.1063/1.3492841,PhysRevB.89.035204,10.1063/1.5047808,PhysRevB.90.235203} and experimental \cite{10.1063/5.0041608,https://doi.org/10.1002/pssb.202200488,PhysRevB.98.125207,Kogiso_2019} findings. 
The CCDs in Fig.~\ref{CCD} illustrate the process of electron or hole capture. 
The diagrams clearly show that for both transition levels the energy barrier for electron capture (transition between green and blue curves) is very large, which leads to low MPE rates.
Energies were evaluated within a small displacement range and used in a parabolic fit;  
the resulting frequencies are shown in the figure.
$\Delta Q$, the separation between equilibrium positions in the two charge states, is 1.632~$\mathrm{amu^{1/2}}${\AA} for (+/0) and 1.695~$\mathrm{amu^{1/2}}${\AA} for (0/$-$). 
The calculated Franck-Condon (FC) energies and HR factors are listed in Sec. S4
of the SM.

Our calculated recombination coefficients (including the Sommerfeld parameters) for the MPE, radiative, and TAAM processes are shown in Tables~\ref{tab:C} and \ref{tab:T}. 
At $T$ = 390~K the calculated Sommerfeld parameters are $s_e^a$ = 7.2, $s_h^a$ = 6.8 for attractive centers and $s_e^r$ = 0.036, $s_h^r$ = 0.042 for repulsive centers. 
A comparison of our calculated capture coefficients with previous calculations~\cite{PhysRevB.90.075202} and with experiments~\cite{reshchikov2023origin,PhysRevB.98.125207,kanegae2018accurate,Kogiso_2019} is included in Sec. S6
of the SM.  
The calculated hole emission coefficient for the (+/0) level at 390~K is $e_p$=$3\times 10^{11}$$\mathrm{s^{-1}}$.

\begin{table}
	\centering
	\caption{Calculated capture coefficients ($\mathrm{cm^3/s}$), at 390~K, for radiative and nonradiative (MPE) recombination  for the (+/0) and (0/$-$) levels of $\mathrm{C_N}$ in GaN.}
        \renewcommand{\arraystretch}{1.35}
	\begin{ruledtabular}
		\begin{tabular}{ccccc}
			Level ($q$/$q-1$) & $C_{p,{\rm MPE}}^{q-1}$ & $C_{p,{\rm rad}}^{q-1}$ & $C_{n,{\rm MPE}}^{q}$ & $C_{n,{\rm rad}}^{q}$ \\ \hline
			$q$=+1 & $4\times10^{-6}$ & $1\times10^{-15}$ & $6\times10^{-19}$ & $9 \times10^{-13}$ \\
			$q$=0 & $1\times10^{-9}$ & $ 1 \times10^{-12}$ & $3\times10^{-21}$ & $7\times10^{-14}$ \\
		\end{tabular}
	\end{ruledtabular}
	\label{tab:C}
\end{table}

\begin{table}
	\centering
	\caption{Calculated TAAM coefficients ($\mathrm{cm^6/s}$) for the (+/0) and (0/$-$) levels of $\mathrm{C_N}$ in GaN.}
       \renewcommand{\arraystretch}{1.35}
	\begin{ruledtabular}
		\begin{tabular}{cccccc}
			Level  ($q$/$q-1$) & $T_1^q$ & $T_2^{q-1}$ & $T_3^{q-1}$ & $T_4^q$ \\ \hline
			$q$=+1 & $2\times 10^{-29}$  & $9\times 10^{-31}$ & $1\times 10^{-31}$  & $3\times 10^{-32}$ \\
			$q$=0 & $2\times 10^{-30}$  & $2\times 10^{-28}$ & $5\times 10^{-31}$  & $6\times 10^{-32}$  \\
		\end{tabular}
	\end{ruledtabular}
	\label{tab:T}
\end{table}


The total density of defects $N_t$ is the sum of defect densities in all charge states: $ N_t = N^-+N^0+N^+$,  
and in steady state, the density of traps in each charge state is constant \cite{PhysRevB.93.201304}.
To simplify the notation, we introduce a generalization of the $C$ coefficients that represent MPE and radiative capture, to also include TAAM and thermal emission.
In the spirit of the $ABC$ model that is commonly used to describe carrier recombination, we additionally assume $n=p$ (usually the case at higher carrier concentrations due to charge neutrality).  The redefined capture coefficients, which we denote with the symbol $K$, are then: 
\begin{align}
    K^0_p & = C_p^0+T_2^0n+T_3^0n \, ,  \\
    K^-_p & = C_p^-+T_2^-n+T_3^-n + e_n/n \, , \\
    K^+_n & = C_n^{+}+T_1^+n+T_4^+n + e_p/n \, , {\rm and}\\
    K^0_n & = C_n^0+ T_1^0n+T_4^0n \, .
\end{align}
Note that these coefficients are now explicitly carrier-density dependent.
This allows us to obtain the total trap-assisted recombination rate $R_{\mathrm{tot}}$:
\begin{equation}
    R_{\mathrm{tot}} = n \, A = n \, N_t \,  K^-_p
    \dfrac{ K^+_n (K^0_p  + K^0_n) - K^0_p e_p/n }
    {K^+_n K^-_p + K^0_p K^-_p + K^0_n K^+_n} 
    \label{Total_R_eq}
\end{equation}
In Sec. S2 
of the SM we offer expressions for the general case where $n$ may be different from $p$.

\begin{figure}[h]
	\centering
	\includegraphics[width=0.85\linewidth]{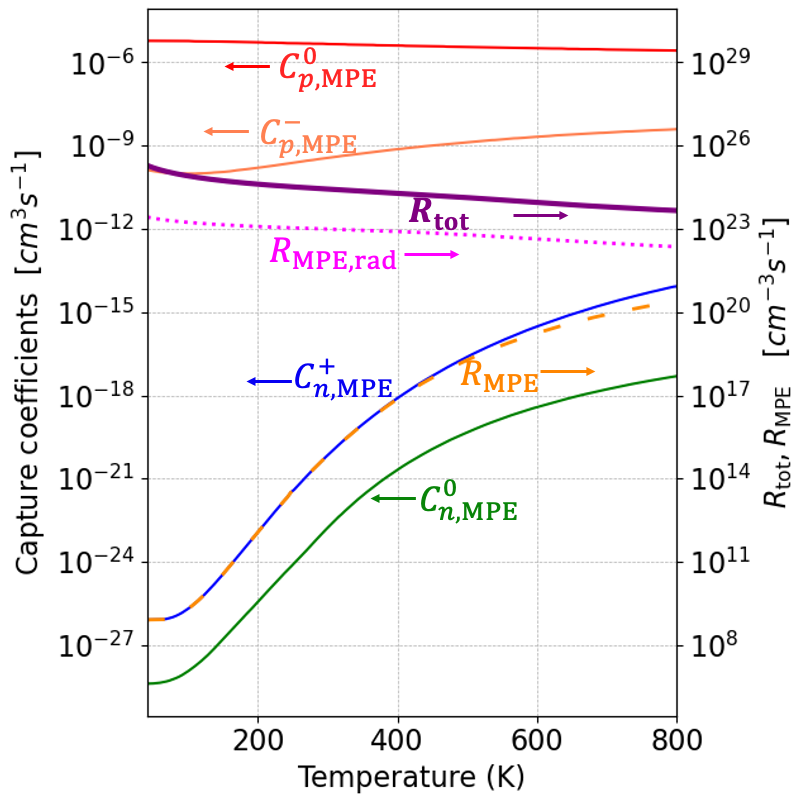}
	\caption{
    Calculated recombination rate as a function of temperature, assuming $N_t = 10^{17}$~cm$^{-3}$ and $n=p = 10^{18}$~cm$^{-3}$.  
    $R_{\rm tot}$ (purple curve) includes the  MPE, radiative processes plus all the  TAAM processes [Eq.~(\ref{Total_R_eq})], while $R_{\rm MPE}$ (red dashed curve) includes only the MPE processes, and $R_{\rm MPE,rad}$ (red dotted curve) includes MPE and radiative processes.
    }
	\label{fig:Rtot}
\end{figure}

In Fig.~\ref{fig:Rtot} we plot the $C$ coefficients and the total recombination rate as a function of temperature,
assuming a concentration $N_t=10^{17}$~cm$^{-3}$ (a typical value for MOCVD-grown GaN \cite{Meyer_2020, fichtenbaum2008impurity,MOE2007710}), and an operating carrier density $n = 10^{18}$ cm $^{-3}$.
The figure confirms that electron capture is the bottleneck in the overall recombination cycle.
We consider several (hypothetical) scenarios.
If we only consider MPE processes, $R_{\mathrm{MPE}}$ virtually coincides with $C_{n,MPE}^+$ (with a suppression at higher temperature due to large hole thermal emission), indicating that nonradiative electron capture at the (+/0) level is the rate-limiting step.
As seen in Table~\ref{tab:C}, the radiative capture rate for electrons is actually much larger than the MPE rate.  If we take both MPE and radiative capture into account, the resulting $R_{\mathrm{MPE,rad}}$ becomes $9 \times10^{22}$~cm$^{-3}$s$^{-1}$ (at 390~K), which is equal to 
$C_{n,{\rm rad}}^{+} N_t n$.  This is still more than two orders of magnitude smaller than the band-to-band radiative recombination rate,~\cite{Dreyer_2016} so would have a negligible impact on device efficiency.

However, if we additionally take TAAM into account, the total trap-assisted recombination rate [Eq.~(\ref{Total_R_eq})] is $R_{\rm tot} \sim 2 \times 10^{24}$~cm$^{-3}$ s$^{-1}$ (at 390~K), 
which is now large enough to impact quantum efficiency. 
TAAM and radiative processes, and hence the resulting $R_{\rm tot}$, depend only weakly on temperature (mainly through the Sommerfeld parameter). 
Above $\sim$600~K increased thermal emission of holes [Eq.~(\ref{e_p})] further suppresses $R_{\rm tot}$.  

TAAM processes also do not depend strongly on the transition energy~\cite{PhysRevLett.131.056402} (again, unlike MPE).
Since the recombination rate for C$_{\rm N}$ in GaN is limited by the TAAM rate for electron capture, we expect this rate (at fixed $N_t$) to remain relatively constant as the bandgap increases.
Thus $\rm C_N$ can also lead to efficiency loss in UV LEDs,~\cite{Kneissl2019} particularly since C incorporation tends to increase with Al content.~\cite{Parish2000}
$R_{\mathrm{tot}}$ is also expected to remain similar in materials with gaps smaller than GaN, up to the point where the gap is small enough for electron capture by MPE to overcome capture by TAAM; this will only be the case in materials with gaps of about 2~eV or smaller.  This would require an In content approaching 40\% in InGaN alloys, a value that well exceeds the values in current devices. 

It is highly informative to examine the recombination rate as a function of injected carrier density, 
allowing us to comment on 
the influence of thermal emission of holes and on the role of radiative capture.
For purposes of this investigation, we assume that the semiconductor has an $n$-type background doping level $n_0$. In the presence of carrier injection or excitation, we then have total free carrier densities $n = n_0 + \Delta n$ and  $p = \Delta p$. 

\begin{figure}[h]
	\centering
	\includegraphics[width=\linewidth]{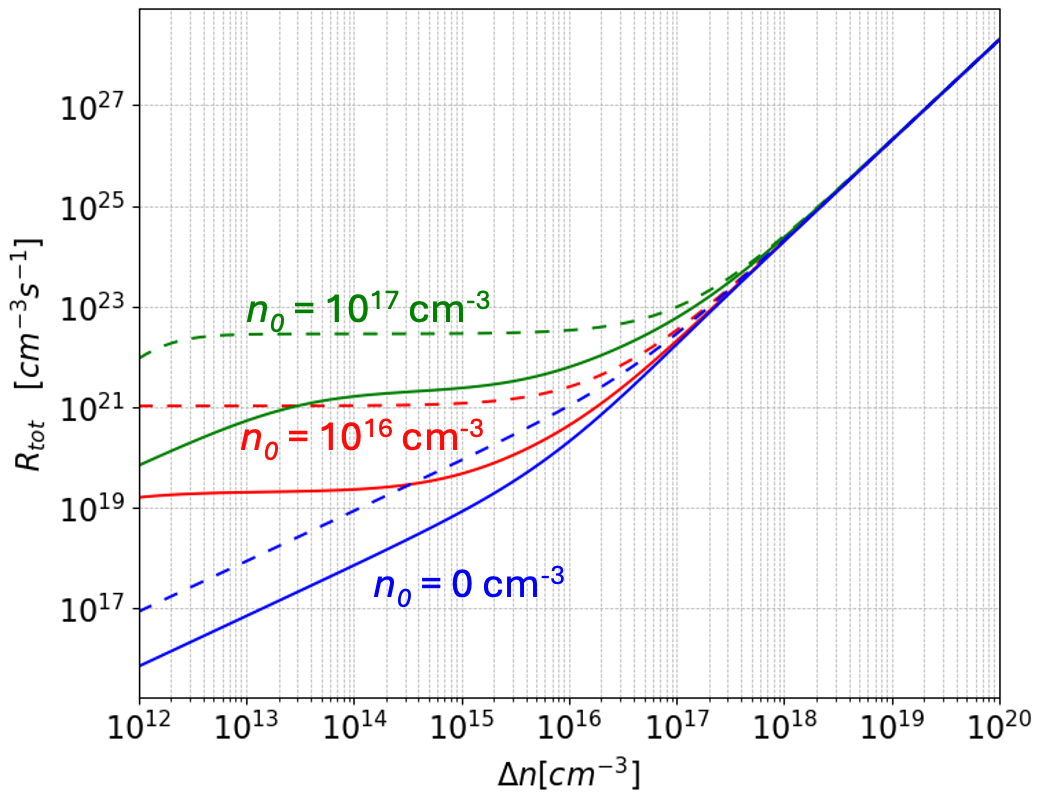}
	\caption{Total recombination rate $R_{\mathrm{tot}}$ for C$_{\rm N}$ in GaN as a function of injected carrier density $\Delta n$, with (solid lines) and without (dashed lines) inclusion of thermal emission (TE) from the (+/0) level. 
    Various levels of background doping $n_0$ ($n_0 = 0, \, \mathrm {10^{16} cm^{-3}}, \, \mathrm {10^{17} cm^{-3}}$) are considered such that $n=n_0+\Delta n$ and $p = \Delta p = \Delta n$. }
	\label{r_tot_fig}
\end{figure}

Figure~\ref{r_tot_fig} shows the calculated $R_{\rm tot}$ [based on Eq.~(S24
) in the SM, which allows for $n \neq p$] as a function of the excited carrier density $\Delta n = \Delta p$. 
Since the rate depends on $n = n_0 + \Delta n$, the presence of background doping leads to a plateau in $R_{\mathrm{tot}}$ at $\Delta n$ values below $n_0$.
The inclusion of thermal emission reduces $R_{\rm tot}$; we attribute this to the fact that fewer defects are available in the positive charge state that is very effective at capturing electrons---see the high value of $T_1^+$ in Table~\ref{tab:T}).
This reduction is more noticeable at lower $\Delta n$ values (see Fig. S3
in the SM), which can be understood as follows.
Inspection of the rate equations for $N^+$ and $N^-$ [Eqs. (S11) and (S12) 
in the SM] shows that 
at high carrier densities hole emission can be neglected, and because hole capture is much faster than electron capture, 
most of the $\rm C_N$ impurities are in the +1 charge state; this is confirmed in Fig. S3
in the SM.
Equation~(S15)
then shows that the total rate is dominated by the TAAM electron capture process into the $(+/0)$ level: 
\begin{equation}
        R_{\rm tot} \approx N^+(T_1^+n+T_4^+p)n \approx N_t T_1^+n^2 \, ,
\label{eq:Rtotapprox1}
\end{equation}
where we also used the fact that $T_1^+ \gg T_4^+$ (Table~\ref{tab:T}).

If we now include thermal emission, Eq. (S11)
in the SM
shows that it will be important if $e_p$ becomes comparable in magnitude to $pC_p^0$.  It therefore makes sense to define a critical hole density $p^*$ such that $p^*C_p^0=e_p$; this corresponds to the density of holes in the VB for the case where the Fermi level coincides with the (+/0) defect level~\cite{SR}.
Using the values in Table~\ref{tab:C} we find $p^*$=7$\times$10$^{16}$~cm$^{-3}$; the curves in Fig.~\ref{r_tot_fig} indeed exhibit an inflection point at that value.

If $\Delta n$ is well above $p^*$, the total rate is given by Eq.~(\ref{eq:Rtotapprox1}). 
On the other hand, when $\Delta n$ is below $p^*$, 
emission out of the (+/0) level will be significant, decreasing $N^+$ and increasing $N^0$ (see Fig. S3
in the SM).
Thus, when $\Delta n \ll p^*$, most of the $\mathrm{C_N}$  are in the neutral charge state, and the total rate [Eq. (S15)]
is dominated by TAAM electron capture into the (0/$-$) level:
\begin{equation}
        R_{\rm tot} \approx N^0(T_1^0 n+T_4^0 p)n \approx N_t T_1^0 n^2\, ,
\label{eq:Rtotapprox2}
\end{equation}
which is a lower rate than in the absence of thermal emission [Eq.~(\ref{eq:Rtotapprox1})] because $T_1^0 < T_1^+$.

The curves in Fig.~\ref{r_tot_fig} actually also contain information about the competition between radiative and TAAM capture of electrons.
If we can assume that the positive charge state $N^+$ dominates, then a comparison between $n C_{n,{\rm rad}}^+$ and $n^2 T_1^+$ indicates that radiative capture will dominate at $n$ below 
5$\times$10$^{16}$~cm$^{-3}$; Fig.~\ref{r_tot_fig} indeed shows a change in the slope of $R_{\rm tot}$ at that value.

For the purpose of analyzing devices such as light-emitting diodes, we are clearly in the regime where $\Delta n \gg p^*$ and hence thermal emission of holes can be ignored.
Figure~\ref{r_tot_fig} does show, however, that unless the injected (or excited) carrier density $\Delta n$ exceeds $n_0$, the presence of background doping will increase $R_{\rm tot}$. Doping of the active layer should thus be avoided since it will adversely impact the efficiency.


In summary, we have scrutinized the impact of carbon impurities on the efficiency of nitride-based light emitters.
We derived the total trap-assisted recombination rate $R_{\rm tot}$ that takes MPE, radiative, TAAM, and thermal emission into account, for a trap with two levels in the bandgap.
While $R_{\rm tot}$ can be formulated in terms of redefined capture coefficients, the resulting SRH $A$ coefficient familiar from the widely used $ABC$ model~\cite{10.1063/1.5096773} [Eq.(\ref{Total_R_eq})] is now carrier-density dependent.
We find that electron capture is the bottleneck in trap-assisted recombination, and that this process becomes dominated by TAAM at carrier densities above 5$\times$10$^{16}$~cm$^{-3}$.
We also find that the thermal emission is unimportant to the total recombination rate when the carrier density is above 7$\times$10$^{16}$~cm$^{-3}$.
Our study highlights the importance of controlling carbon concentrations in GaN-based devices, and highlights the crucial role  of TAAM processes in trap-assisted recombination in wide-bandgap semiconductors.

See the Supplementary Material for details on the calculation of TAAM coefficients, a derivation of the total recombination rate, details of first-principles calculations and the formation-energy diagram, quantities relevant for the CCD, and trap concentrations as a function of carrier density.

We acknowledge 
S. Keller for fruitful discussions. 
F. Z., H. G., and M. E. T. were supported by the U.S. Department of Energy (DOE), Office of Science, Basic Energy Sciences (BES) under Award No. DE-SC0010689. 
F. Z. also acknowledges support from the California NanoSystems Institute for an Elings Prize Fellowship, and the Alexander von Humboldt-Stiftung for the support from the Humboldt Research Fellowship. 
C. G. VdW. was supported by the Office of Naval Research, Award No. N00014-22-1-2808 (Vannevar Bush Faculty Fellowship). 
We acknowledge the use of resources of the National Energy Research Scientific Computing Center, a DOE Office of Science
User Facility supported by the Office of Science of the U.S. Department of Energy under Contract No. DE-AC02-05CH11231 using NERSC Award No. BES-ERCAP0021024.

\section*{Author declarations}

\subsection*{Conflict of Interest}

The authors have no conflicts to disclose.

\subsection*{Author Contributions}

F. Zhao: Conceptualization (equal); Formal analysis (equal); Investigation (equal); Writing - original draft (equal); Writing - review \& editing (equal).
H. Guan: Formal analysis (equal); Investigation (equal); Writing - original draft (equal); Writing - review \& editing (supporting).
M.\ E.\ Turiansky: Conceptualization (equal); Formal analysis (supporting); Investigation (supporting); Writing - review \& editing (supporting).
C. G. Van de Walle: Conceptualization (equal); Formal analysis (equal); Writing - review \& editing (equal); Supervision (lead); Funding acquisition (lead).

\section*{DATA AVAILABILITY}

The data that support the findings of this study are available from the corresponding author upon reasonable request.

\bibliographystyle{apsrev4-2}
\bibliography{main.bib}

\end{document}


\pagestyle{fancy}
\title{Supplementary Material for \\ Carbon in GaN as a nonradiative recombination center}
\author{Fangzhou Zhao}
\affiliation{Materials Department, University of California, Santa Barbara, California 93106-5050, USA 
}
\affiliation{Theory Department, Max Planck Institute for Structure and Dynamics of Matter and Center for Free-Electron Laser Science, 22761 Hamburg, Germany}
\date{\today}
\author{Hongyi Guan}
\affiliation{Materials Department, University of California, Santa Barbara, California 93106-5050, USA 
}
\author{Mark E. Turiansky}
\altaffiliation[Present Address: ]{US Naval Research Laboratory, 4555 Overlook Avenue SW, Washington, DC 20375, USA}
\affiliation{Materials Department, University of California, Santa Barbara, California 93106-5050, USA 
}
\author{Chris G. Van de Walle}
\affiliation{Materials Department, University of California, Santa Barbara, California 93106-5050, USA 
}
\maketitle

\section{Calculation details of trap-assisted Auger-Meitner recombination coefficient}
\label{Ssec:TAAM}

Based on Ref.~\onlinecite{PhysRevLett.131.056402}, the four trap-assisted Auger-Meitner (TAAM) coefficients are
\begin{eqnarray}
    T_1 &&=  \frac{2\pi}{ \hbar} V^2 \sum_{ \bm 4 \in {\rm CB}} | M^1_{\bm 1\bm 2\bm t\bm 4} |^2  \delta (\epsilon_{\bm 1}+ \epsilon_{\bm 2} - \epsilon_{\bm t} - \epsilon_{\bm 4}), \\
    T_2 &&=  \frac{2\pi}{ \hbar} V^2 \sum_{ \bm 4 \in {\rm VB}} | M^2_{\bm 1\bm 2\bm t\bm 4} |^2  \delta (\epsilon_{\bm 1}+ \epsilon_{\bm 2} - \epsilon_{\bm t} - \epsilon_{\bm 4}), \\
    T_3 &&=  \frac{2\pi}{ \hbar} V^2 \sum_{ \bm 4 \in {\rm CB}} | M^3_{\bm 1\bm 2\bm t\bm 4} |^2  \delta (\epsilon_{\bm 1}+ \epsilon_{\bm t} - \epsilon_{\bm 2} - \epsilon_{\bm 4}), \\
    T_4 &&=  \frac{2\pi}{ \hbar} V^2 \sum_{ \bm 4 \in {\rm VB}} | M^4_{\bm 1\bm 2\bm t\bm 4} |^2  \delta (\epsilon_{\bm 1}+ \epsilon_{\bm 4} - \epsilon_{\bm t} - \epsilon_{\bm 2}),   \label{TAAM}
\end{eqnarray}
where CB refers to bulk states in the conduction band and VB to bulk states in the valence band, and $t$ denotes the trap level. 
The matrix elements include the direct and exchange kernels of the screened Coulomb interaction $\hat{W}$; for example:
\begin{eqnarray}
 M^1_{\bm 1\bm 2\bm t\bm 4} =\bra{\bm 1\bm  2} \hat{W} \ket{\bm t \bm 4} - \bra{\bm 1\bm  2} \hat{W} \ket{\bm 4 \bm t}. 
\end{eqnarray}
$\hat{W}$ contains a model dielectric function \cite{cappellini1993model}. 
The above equations slightly differ from those in Ref.~\onlinecite{PhysRevLett.131.056402} because we take into account that only free carriers at the band edges contribute to the recombination process and the Fermi-Dirac distribution function can be neglected.

\begin{figure}
	\centering
	\includegraphics[width=\linewidth]{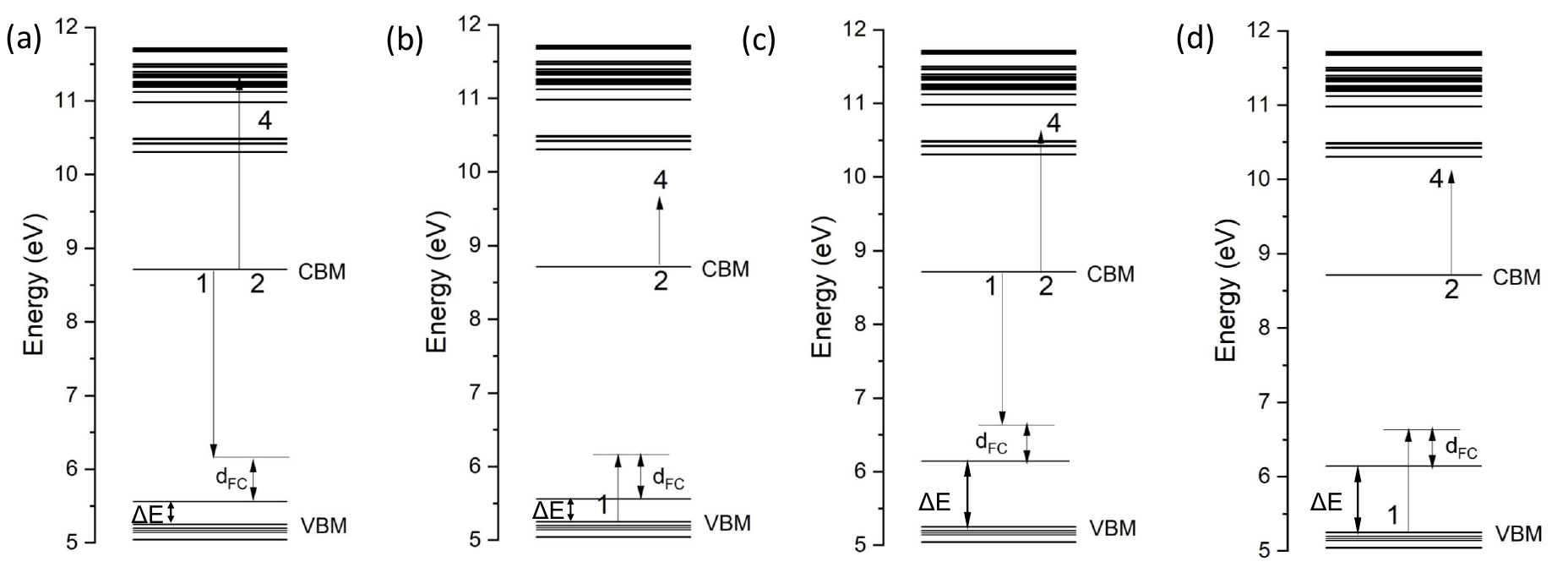}
	\caption{Kohn-Sham states relevant for calculation of the (a) $T_1^+$ and (b) $T_3^+$ coefficients for the (+/0) level, and (c) $T_1^0$ and (d) $T_3^0$ coefficients for the (0/$-$) level. 
    CBM denotes the conduction-band minimum, VBM the valence-band maximum.
    Energy conservation is indicated by the length of upper and lower arrows being equal.}
	\label{EN_diag}
\end{figure}

The evaluation of TAAM coefficients is based on wavefunctions calculated in a supercell. However, when the  CB or VB is very dispersive, bulk states are only sparsely sampled in a supercell. 
In GaN, this is particularly a problem for the CB, affecting the evaluation of $T_1$ and $T_3$.
We overcome this issue by replacing the summation over the bulk CB states by a continuous integration according to the CB density of states (DOS) $D(\epsilon)$:
\begin{equation}
    T_1 = \dfrac{2\pi}{\hbar} V^2 \int_{CBM}^{\infty} d\epsilon_{\bm 4}  D(\epsilon_{\bm 4})  \times \overline{|M^1_{\bm 1\bm 2\bm t \bm 4}|^2}  \delta (\epsilon_{\bm 1}+\epsilon_{\bm 2}-\epsilon_{\bm t}-\epsilon_{\bm 4})  ,
\end{equation}
where $\overline{|M^1_{\bm 1\bm 2\bm t \bm 4}|^2}$ refers to an average over the $\bm 4^{th}$ state in the CB~\cite{PhysRevLett.131.056402}. 

Figure~\ref{EN_diag} shows the Kohn-Sham states at the $\Gamma$ point in a 360-atom supercell relevant for the $T_1$ and $T_3$ coefficients of the (+/0) and (0/$-$) levels.
In Fig.~\ref{EN_diag}(a), the energy level of the $4^{th}$ state falls within a range of CB energies that are well sampled in the supercell, so $T_1^+$ could be calculated using either the direct summation method or the DOS integration method. However, in Figs.~\ref{EN_diag}(b)-(d), the bulk states near the energy-conserving transitions are clearly not well sampled, necessitating the use of the DOS integration method.

We also take phonon assistance in the TAAM recombination into account. 
For C$_{\rm N}$ in GaN the electron phonon coupling is strong (large Huang-Rhys factors), so the electron phonon spectral function can be approximated by a Gaussian function.  
The broadening parameter $\sigma_e$ ($\sigma_h$) for the electron (hole) capture process is estimated based on the 1D configuration coordinate diagram \cite{jia2020design, alkauskas2016tutorial}: 
 \begin{eqnarray}
\sigma_e &&=  \hbar \Omega_e \sqrt{S^e } \sqrt{\cot \left(\frac{\hbar\Omega_e}{2k_B T} \right)}  \\
\sigma_h &&=  \hbar \Omega_g \sqrt{S^h } \sqrt{\cot \left(\frac{\hbar\Omega_g}{2k_B T} \right)} \, .
 \label{eq:sigma}
 \end{eqnarray}
$\Omega_e$ and $\Omega_g$ are the effective phonon frequency of the $-$1 charge state and the neutral charge state, respectively, in a 1D harmonic approximation. 
We find that both $\sigma_e$ and $\sigma_h$ are equal to 0.22 eV.
We also replace the Kohn-Sham trap level $\epsilon_t$ by the thermodynamic transition energy $\Delta E$ plus the Frank-Condon energy of the ground state $d_{FC}$ \cite{PhysRevLett.131.056402}:
\begin{equation}
    \delta (\varepsilon_{1}+\varepsilon_{2}-\varepsilon_{t}-\varepsilon_{4}) \approx g[\varepsilon_{1}+\varepsilon_{2}-\Delta E-\varepsilon_{4}-d_{FC}]
\end{equation}

\section{Derivation of the total recombination rate}
\label{Ssec:Rtot}

In this section, we derive the total recombination rate of the $\mathrm{C_N}$ impurity in GaN, which has two thermodynamic transition levels: (+/0) and (0/$-$). We include the effects of radiative and nonradiative [multiphonon emission (MPE)] capture, TAAM recombination, as well as thermal emission from the trap level.

The nonradiative MPE rate~\cite{PhysRevB.90.075202} and the radiative recombination rate~\cite{dreyer2020radiative} depend on capture coefficients in the same way.
We can thus define the coefficients at a $(q/ q-1)$ level as follows: for nonradiative MPE recombination, we use $C^q_{n,{\rm MPE}}$ for electron capture and $C^q_{p,{\rm MPE}}$ for hole capture; for radiative recombination we use $C^q_{n,{\rm rad}}$ for electrons and $C^q_{p,{\rm rad}}$ for holes. 
For purposes of writing rate equations, we can lump MPE and radiative recombination together and define $C^q_{n} = C^q_{n,{\rm MPE}} + C^q_{n,{\rm rad}}$ and $C^q_{p} = C^q_{p,{\rm MPE}} + C^q_{p,{\rm rad}}$.

We can write rate equations for electron and hole densities, and set them to zero to reflect steady-state conditions~\cite{Abakumov1991,PhysRevB.64.115205}:
\begin{equation}
    \dfrac{dn}{dt} = E_n-R_n+G=0, \quad \dfrac{dp}{dt} = E_p-R_p+G=0, 
    \label{eq:rates}
\end{equation}
where $R_n$ and $R_p$ are the recombination rates of electrons and holes, $G$ (units $\mathrm{cm^{-3}s^{-1}}$) the generation rate of electron-hole pairs,
and $E_n= N^- e_n$ and $E_p= N^+e_p$ the thermal emission rates. 
In steady state, the density of traps in each charge state is also constant \cite{PhysRevB.93.201304}:
\begin{align}
&
\begin{aligned}
& \frac{dN^+}{dt} = - (nC_n^{+}+T_1^+n^2+T_4^+np+e_p)N^+ + (C_p^0p+T_2^0p^2+T_3^0np)N^0 = 0 
\end{aligned} \label{+/0_equil} \\
&
\begin{aligned}
& \frac{dN^-}{dt} = (nC_n^0+T_1^0n^2+T_4^0np)N^0 - (pC_p^-+T_2^-p^2+T_3^-np  + e_n)N^-  = 0
\end{aligned} \label{0/-_equil} 
\end{align}
The total density of defects $N_t$ is the sum of defect densities in all charge states:
\begin{equation}
    N_t = N^-+N^0+N^+ \label{defect}
\end{equation}
The total effective recombination rate is the recombination rate minus the thermal emission rate \cite{Abakumov1991}:
\begin{equation}
    R_{\mathrm{tot}} = R_n-E_n = R_p-E_p = G \, .
    \label{rtot}
\end{equation} 
The total hole recombination rate is the sum of all hole recombination rate terms \cite{PhysRevB.93.201304}:
\begin{equation}
\begin{aligned}
     R_p &=N^0(C_p^0p+T_2^0p^2+T_3^0np)+N^-(pC_p^-+T_2^-p^2+T_3^-np + e_n) \\
   &= N^+(nC_n^{+}+T_1^+n^2+T_4^+np+e_p] + N^0(nC_n^0+ T_1^0n^2+T_4^0 np) \, .
\end{aligned} \label{r_p}
\end{equation}
With Eqs.~(\ref{+/0_equil}) and (\ref{0/-_equil}), we can get the ratio between $N^-$, $N^0$ and $N^+$. 
Plugging all of this into Eq.~(\ref{rtot}) we finally obtain the total trap-assisted recombination rate $R_{\mathrm{tot}}$:
\begin{equation}
    \resizebox{\textwidth}{!}{$
    R_{\mathrm{tot}} = N_t 
    \dfrac{ (pC_p^-+T_2^-p^2+T_3^-np + e_n)
    \Big((C_p^0p+T_2^0p^2+T_3^0np )[nC_n^{+}+(T_1^+n+T_4^+p)n]
    +(nC_n^0+T_1^0n^2+T_4^0np)[nC_n^{+}+(T_1^+n+T_4^+p)n+e_p]\Big)}{(pC_p^-+T_2^-p^2+T_3^-np + e_n)
    [nC_n^{+}+(T_1^+n+T_4^+p)n+e_p+C_p^0p+T_2^0p^2+T_3^0np]+ 
    (nC_n^0+T_1^0n^2+T_4^0np)
    [nC_n^{+}+(T_1^+n+T_4^+p)n+e_p]}
    $}  
    \label{Total_R_eq}
\end{equation}

We note that in this expression the $C$ coefficients (for MPE and radiative recombination) appear on equal footing as $T$ coefficients (for TAAM) multiplied by the carrier density, consistent with TAAM becoming more important at higher carrier densities. 
In the same vein, $C_n^+$ and $e_p/n$ appear on the same footing, reflecting the similarity of electron capture and hole emission, and indicating that thermal emission of holes will become relatively {\it less} important at higher electron densities.
The same applies to  $C_p^-$ and $e_n/p$, reflecting the similarity of electron capture and hole emission, and indicating that thermal emission of electrons will become relatively less important at higher hole densities.

These observations also inspire us to simplify Eq.~(\ref{Total_R_eq}) by introducing capture coefficients (which we will denote with the symbol $K$)
that include not only MPE and radiative capture, but also TAAM and thermal emission.
This will allow us to derive an $A$ coefficient (representing Shockley-Read-Hall recombination) in the spirit of the widely used $ABC$ model for carrier recombination and internal quantum efficiency, with a dependence of $A$ on capture coefficients that will look similar to conventional expressions; the difference will be that the capture coefficients are now carrier-density dependent.
The $ABC$ model also makes the assumption that $n$=$p$, but for now we keep things general and allow electron and hole densities to be different. We introduce coefficients $\tilde{K}$ defined as:
\begin{align}
\begin{aligned}    
&    \tilde{K}^0_p = C_p^0+T_2^0p+T_3^0n \, , \\ 
&    \tilde{K}^-_p = C_p^-+T_2^-p+T_3^-n + e_n/p \, , \\   
&    \tilde{K}^+_n = C_n^{+}+T_1^+n+T_4^+p + e_p/n \, , {\rm and} \\  
&    \tilde{K}^0_n = C_n^0+ T_1^0n+T_4^0p \, .
\end{aligned} 
\label{Atilde} 
\end{align}

Using these coefficients, we can rewrite Eqs.~(\ref{+/0_equil}) and (\ref{0/-_equil}) as
\begin{align}
&
\begin{aligned}
& \frac{dN^+}{dt} = - n \tilde{K}^+_n N^+ + p \tilde{K}^0_p N^0 = 0 \, ,
\end{aligned} \label{+/0_equil_tilde} \\
&
\begin{aligned}
& \frac{dN^-}{dt} = n \tilde{K}^0_n N^0 - p \tilde{K}^-_p N^-  = 0 \, .
\end{aligned} \label{0/-_equil_tilde} 
\end{align}
These steady-state conditions lead to:
\begin{align}
\frac{N^+}{N^0} &= \frac{p}{n}\frac{\tilde{K}_p^0}{\tilde{K}_n^+}  \, , {\rm and} \label{eq:N+tilde}\\
\frac{N^-}{N^0} &= \frac{n}{p}\frac{\tilde{K}_n^0}{\tilde{K}_p^-}  \label{eq:N-tilde}\, ,
\end{align}
Combining these equations with Eq.~(\ref{defect}) we can then solve for the density of neutral traps: 
\begin{equation}
N_0 = \frac{N_t} {1 + \frac{p}{n}\frac{\tilde{K}_p^0}{\tilde{K}_n^+} + \frac{n}{p}\frac{\tilde{K}_n^0}{\tilde{K}_p^-}}  \, .
\label{eq:N0tilde}
\end{equation}
and Eq.~(\ref{r_p}) can be written as
\begin{equation}
\begin{aligned}
   R_p &= N^0 p \tilde{K}^0_p + N^- p \tilde{K}^-_p\\
   &= N^+ n \tilde{K}^+_n + N^0 n \tilde{K}^0_n \, .
\end{aligned} \label{r_p_tilde}
\end{equation}
Equation~(\ref{Total_R_eq}) can then be written as
\begin{equation}
\begin{aligned}
      R_{\mathrm{tot}} & = 
    \dfrac{ 
    p N_t \tilde{K}^-_p 
    [ p \tilde{K}^0_p n (\tilde{K}^+_n - e_p/n)
    + n \tilde{K}^0_n n \tilde{K}^+_n ] }
    { p \tilde{K}^-_p 
   ( n \tilde{K}^+_n + p \tilde{K}^0_p)+ 
    n \tilde{K}^0_n 
     n \tilde{K}^+_n} \\
     &=
    p N_t \tilde{K}^-_p 
     \dfrac{ n \tilde{K}^+_n (p \tilde{K}^0_p + n \tilde{K}^0_n) - p \tilde{K}^0_p  e_p }
    {  n \tilde{K}^+_n p \tilde{K}^-_p +  p \tilde{K}^0_p p \tilde{K}^-_p  + n \tilde{K}^0_n  n \tilde{K}^+_n}
\end{aligned}
    \label{Total_R_eq_tilde}
\end{equation}

The derivations and results so far address the general case where the electron density $n$ and hole density $p$ can be different.
In the $ABC$ model, it is commonly assumed (based on local charge neutrality) that $n$=$p$.
The expressions for the $K$ coefficients then simplify to:
\begin{align}
\begin{aligned}    
&    K^0_p = C_p^0+T_2^0n + T_3^0n \, , \\ 
&    K^-_p = C_p^-+T_2^-n+T_3^-n + e_n/n \, , \\ 
&    K^+_n = C_n^{+}+T_1^+n+T_4^+n + e_p/n \, , {\rm and} \\
&    K^0_n = C_n^0+ T_1^0n+T_4^0n \, .
\end{aligned} 
\label{A} 
\end{align}
The density of the neutral traps [Eq.~(\ref{eq:N0tilde})] becomes
\begin{equation}
N_0 = \frac{N_t} {1 + \frac{K_p^0}{K_n^+} + \frac{K_n^0}{K_p^-}}  \, ,
\end{equation}
and the hole capture rate [Eq.~(\ref{r_p_tilde})] can be expressed as:
\begin{align}
\begin{aligned}
   R_p &= n (N^0 K^0_p + N^- K^-_p)\\
   &= n (N^+ K^+_n + N^0 K^0_n)\, .
\end{aligned} \label{r_p_np}
\end{align}
Finally, the total recombination [Eq.~(\ref{Total_R_eq_tilde})], assuming $n$=$p$, becomes:
\begin{equation}
        R_{\mathrm{tot}} = n \,  N_t K^-_p
    \dfrac{ K^+_n (K^0_p  + K^0_n) - K^0_p e_p/n }
    {K^+_n K^-_p + K^0_p K^-_p + K^0_n K^+_n} 
    \label{Total_R_eq_sim}
\end{equation}

\section{Details of first-principles calculations and formation energy}
\label{Ssec:formE}

We use norm-conserving pseudopotentials~\cite{fuchs1999ab} with Ga $3d$ in the core; tests showed that including the $d$ states in the valence did not affect the conclusions. 
With a mixing parameter $\alpha = 0.33$ and a plane-wave cutoff of 120 Ry we obtain lattice parameters $a = 3.19$~{\AA} and $c = 5.18$~{\AA}, and a bandgap of 3.55 eV. These results are consistent with both experimental data \cite{madelung2004semiconductors} and prior computational findings \cite{PhysRevLett.131.056402, PhysRevLett.109.267401,PhysRevB.89.035204}. 

\begin{figure}[H]
	\centering
	\includegraphics[width=100mm]{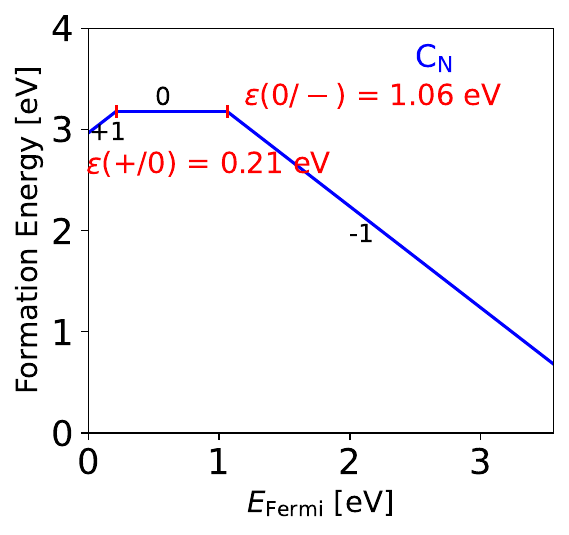}
	\caption{Calculated formation energies for the $\mathrm{C_N}$ impurity in GaN. Results are from a 96-atom supercell, under Ga-rich conditions, and with the C chemical potential determined by bulk diamond. }
	\label{sup-Formation_energy}
\end{figure}

The calculated formation energy diagram is shown in Fig.~\ref{sup-Formation_energy}.

\section{Calculated Franck-Condon energies and Huang-Rhys factors}
\label{Ssec:FCHR}

\begin{table}[H]
	\centering
	\caption{Calculated Franck-Condon (FC) energies $d_{\mathrm{FC}}^{h}$ and $d_{\mathrm{FC}}^{e}$ and Huang-Rhys factors $S^h$ and $S^e$.}
        \renewcommand{\arraystretch}{1.35}
	\begin{ruledtabular}
		\begin{tabular}{cccccc}
			Level & $d_{\mathrm{FC}}^{h} $ & $d_{\mathrm{FC}}^{e}$  & $S^h$  & $S^e$  \\ \hline
			(+/0)  &  0.33~eV &  0.60~eV  &   $d_{\mathrm{FC}}^{h} / \Omega_g|_{+} \approx 10 $ & $d_{\mathrm{FC}}^{e} / \Omega_e|_{+} \approx  14 $  \\
			(0/$-$)  &  0.38~eV  & 0.49~eV  & $ d_{\mathrm{FC}}^{h} / \Omega_g|_{0} \approx  11 $ & $ d_{\mathrm{FC}}^{e} / \Omega_g|_{0} \approx 13 $  \\
		\end{tabular}
	\end{ruledtabular}
	\label{summary}
\end{table}

\section{Trap concentrations as a function of carrier density}
\label{Ssec:Chargestate}

\begin{figure}[H]
	\centering
	\includegraphics[width=160mm]{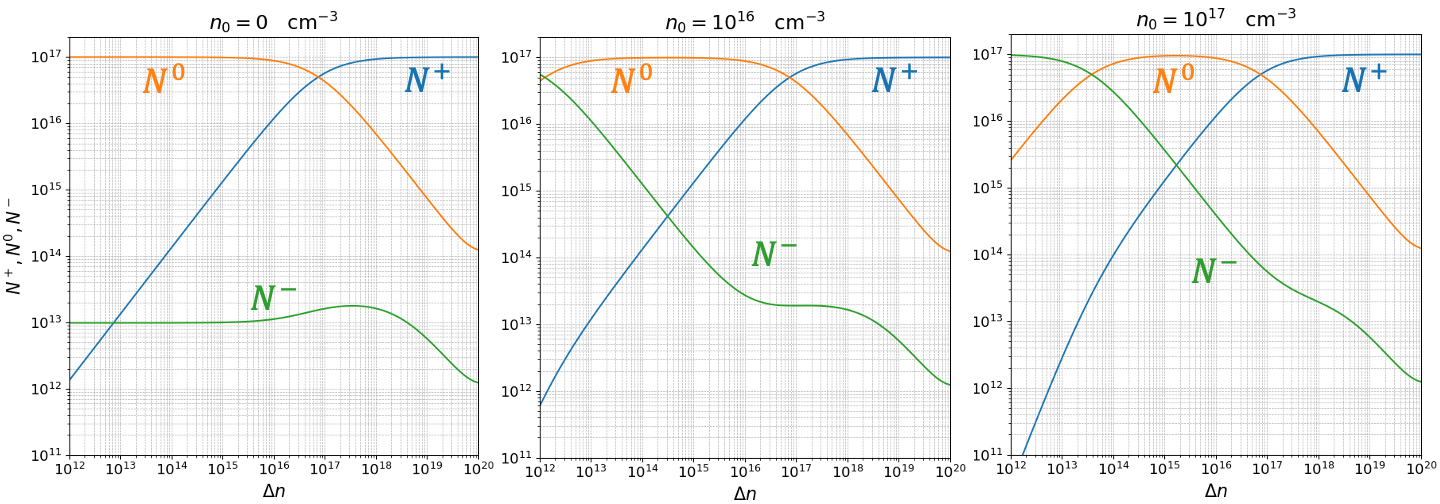}
	\caption{Calculated density of C$_{\rm N}$ traps in different charge states ($N^+$, $N^0$, and $N^-$) as a function of injected carrier density $\Delta n$ for three background doping levels $n_0$. }
	\label{sup-trap_density}
\end{figure}


In Fig.~\ref{sup-trap_density} we show the calculated density of  C$_{\rm N}$ in GaN in different charge states $N^+$, $N^0$, and $N^-$, as a function of the injected carrier density $\Delta n$,
obtained from Eqs.~(\ref{Atilde}) and (\ref{eq:N+tilde})--(\ref{eq:N0tilde}) (allowing for $n$ being different from $p$).
Three levels of $n$-type background doping $n_0$ are considered; the total carrier densities are $n = n_0 + \Delta n$ and $p = \Delta p$. 
MPE and radiative recombination as well as TAAM recombination are included. 
At high injected carrier densities, most of the $\rm C_N$ impurities are in the +1 charge state, since hole capture at the (+/0) level is the dominant process. 
At lower injected carrier densities, thermal emission becomes important.
Equation~(\ref{+/0_equil}) shows that if $e_p$ becomes comparable in magnitude to $pC_p^0$, $N^0$ will be of comparable magnitude as $N^+$; that is why we defined a critical hole density $p^*$ such that $p^*C_p^0=e_p$.
With the values in Table I 
of the main text, $p^*=7 \times 10^{16} \mathrm{cm^{-3}}$, which is indeed the point in Fig.~\ref{sup-trap_density} where $N^0 \approx N^+$.

For $\Delta n \ll p^*$, most of the $\mathrm{C_N}$  are in the neutral charge state, but we see that if the background doping density is significant ($n_0$=10$^{16}$ or 10$^{17}$ cm$^{-3}$ in Fig.~\ref{sup-trap_density}), $N^-$ becomes comparable to or larger at low injected carrier densities; this occurs because in Eq.~(\ref{0/-_equil}) the $n^2 T_1^0$ term becomes larger than the $pC_p^-$ term, and is responsible for the downturn in $R_{\rm tot}$ at $\Delta n$$\sim$2$\times$10$^{13}$~cm$^{-3}$ in Fig. 4 
of the main text.


\section{Comparison of calculated capture coefficients with experiment}
\label{Ssec:Compare_exp}


The calculated electron capture coefficient $C^0_n$ (dominated by $C^0_{n,{\rm rad}}$) matches well with the  experimental value $1.1 \times 10^{-13}$ cm$^{3}$/s reported in Ref.~\onlinecite{PhysRevB.98.125207} for the yellow luminescence (YL1) line. 

The calculated hole capture coefficient $C^-_p$ (dominated by $C^-_{p,{\rm MPE}}$) matches well with previous calculations~\cite{PhysRevB.90.075202}, but
is somewhat smaller than reported experimental values~\cite{kanegae2018accurate, tokuda2016dlts, polyakov2016electrical, polyakov2016deep, reshchikov2023origin,Kogiso_2019}.
Most of the underestimation can be attributed to our use of a zero-temperature value for the (0/$-$) level; at finite temperature this level moves closer to the VBM~\cite{10.1063/1.5047808}, bringing it closer to the experimental transition energy \cite{reshchikov2023origin} and enhancing the hole capture rate~\cite{PhysRevB.90.075202}. 
We emphasize that this underestimate in $C^-_p$ does not affect any of our subsequent conclusions, since hole capture is never the limiting process in the overall recombination cycle. 

For the (+/0) level, Ref.~\onlinecite{Kogiso_2019} reported $C_p$$\sim$1$\times$10$^{-7}$~cm$^{3}$/s, in quite good agreement with our value of $C_{p,{\rm MPE}}^0$=4$\times$10$^{-6}$~cm$^{3}$/s.
Reference~\onlinecite{PhysRevB.98.125207}, on the other hand reported $C_n$$\sim$1$\times$10$^{-9}$~cm$^{3}$/s and $C_p$$\sim$1$\times$10$^{-10}$~cm$^{3}$/s for the (+/0) level.
This $C_p$ value is significantly smaller than our calculated $C^0_{p,{\rm MPE}}$ value, and much smaller than we would expect for a level so close to the VBM; we speculate that significant hole emission may have affected the measured value.
The measured $C_n$$\sim$1$\times$10$^{-9}$~cm$^{3}$/s value, on the other hand, is much larger than our calculated value for the radiative rate; and indeed several orders of magnitude larger than expected rates for radiative transitions~\cite{dreyer2020radiative}.
We speculate that the background electron concentration in the samples of Ref.~\onlinecite{PhysRevB.98.125207} (between 5$\times$10$^{16}$~cm$^{3}$/s and 5$\times$10$^{18}$~cm$^{-3}$/s) may have impacted the extracted value due to the participation of TAAM processes.

\bibliographystyle{apsrev4-2}
\bibliography{main.bib}